# Non-adiabatic coupling as friction in the formation of $H_3^+$: A classical mechanical study

Michael Baer,[1] Soumya Mukherjee,[2] Satyam Ravi,[2,3] Satrajit Adhikari,[2] Narayanasami Sathyamurthy,[4]

[1)] The Fritz Haber Center for Molecular Dynamics, The Hebrew University of Jerusalem, Jerusalem, Israel.

[2)] School of Chemical Sciences, Indian Association for the Cultivation of Science, Kolkata, India.

[3)] School of Advance Science and Languages, VIT Bhopal University, Bhopal, India

[4)] Department of Chemical Sciences, Indian Institute of Science Education and Research Mohali, Sector 81, SAS Nagar, Manauli 140306 India

**Abstract**

By going beyond the Born-Oppenheimer approximation and treating the non-adiabatic coupling terms (NACTs) as equivalent to a frictional force in a molecular system, the classical equations of motion are solved for a test case of $H_3^+$. Using an *ab initio* potential energy surface for the ground electronic state and its NACTs with the first excited state of $H_3^+$, it is shown that ($D^+$, $H_2$) collisions are slowed enough to result in trapping and formation of a stable $DH_2^+$.

# 1. Introduction

Any chemical dynamical problem, in principle, could be solved by solving the time-independent or time-dependent Schrödinger equation for the nuclei and the electrons contained in the system. In practice, however, this is a formidable challenge and has been achieved for very few systems. One usually resorts to the Born-Oppenheimer (BO) approximation for computing (and interpreting the experimentally measured) differential/integral reaction cross sections and rate coefficients for many elementary chemical reactions in a limited range of energy and temperature. When there is more than one potential energy surface becoming degenerate and/or (avoid) crossing each other, the BO approximation fails. A limited number of systems have been investigated quantum mechanically by including the non-adiabatic coupling between potential energy surfaces explicitly [For example, see ref. 1-8]. While the quasi-classical trajectory method[9] has been successful in describing the dynamics on a single (adiabatic) potential energy surface for many systems, trajectory surface hopping and a variety of semiclassical and related methods[10-20] have been used to consider switching between surfaces.

The limitation of the BO approximation and the need to go beyond it have been articulated elsewhere[21,22] and for the present study the Born-Oppenheimer-Huang (BOH)[21-25] non-adiabatic coupling terms (NACTs) $\tau_{Q_k jj+1}$ between two adjacent electronic states $j$ and $j+1$ are defined as:

$$\tau_{Q_k jj+1} = \left\langle \zeta_j \middle| \frac{\partial}{\partial Q_k} \zeta_{j+1} \right\rangle, \quad Q_k = x_k, y_k, z_k ,  \qquad (1)$$

where ($x_k, y_k, z_k$), $k = 1, 2, 3$ are the nuclear (Cartesian) coordinates for a triatomic system.

The NACTs are known to become **singular** at certain points, referred to as Conical Intersections (*ci*s),[21a] in the molecular plane for molecules containing at least three atoms. The number of *ci*s between different electronic states is determined by various parameters that define the system. However, it can be said that systems made up of three identical atoms/ions like (*X-X-X*), (*X-X-X*)$^+$, etc. are likely to have their *ci*s to be located at the *equilateral* positions although often we find them also at the *isosceles* positions.

Because of the form of the BOH equation, namely: [21b]

$$-\frac{\hbar^2}{2\mu}(\nabla + \boldsymbol{\tau}(\mathbf{s}))^2 \Psi(\mathbf{s}) + \mathbf{u}(\mathbf{s})\Psi(\mathbf{s}) = E\Psi(\mathbf{s}) \tag{2}$$

we had suggested recently[26,27] that we could attribute to the NACTs dissipative features and that they could behave like **Friction Forces**. It must be pointed out that Eq. (2) is a **matrix** equation where **u(s)** is a diagonal matrix containing the adiabatic potential energy surfaces (PESs) and **τ(s)** stands for the anti-symmetric NACT matrix (see Eq. (1)), with **s** referring to the nuclear coordinates and $\Psi, E, \nabla, \hbar,$ and $\mu$ have their usual meaning. [21b]

Influence of friction on reaction dynamics in condensed phases has been studied for several decades now.[28] There has been a renewed interest in molecular friction in the context of protein (mis)folding in biological systems[29]. However, one seldom talks about friction in isolated (gas phase) molecular events. In this communication, we point out how friction could be envisaged as arising from non-adiabatic coupling in molecular systems and result in the formation of $DH_2^+$, for example.

## 2. Theoretical Background

The basic classical equation of motion (in one dimension), which contains the friction term, $\beta \dot{x}$, takes the form[30,31]

$$m\ddot{x} + \beta \dot{x} + \frac{dV(x)}{dx} = 0 \tag{3}$$

where the terms $\dot{x}$ and $\ddot{x}$ stand for $(dx/dt)$ and $(d^2x/dt^2)$, respectively and $V$ is the potential energy. This can be modified to

$$\left[\frac{1}{2m}(m\dot{x} + \beta x)^2 + V(x)\right] = G(x, \dot{x}), \tag{4}$$

where

$$G(x, \dot{x}) = \int x \left[-\frac{1}{m}\beta \frac{dV}{dx} + \frac{d\beta}{dt}\left(\frac{1}{m}\beta x + \dot{x}\right)\right] dt. \tag{5}$$

The similarity between the quantum mechanical equation (2) and the classical equation of motion (4) is not obvious since the former is a matrix-equation that must be applied to a group of coupled states whereas the latter is an equation for a single isolated state. Still such a similarity

can be envisaged by applying the modified BO approximation as was derived by Baer and Englman.[32, 21(c)]

This approximation yields for a system of two coupled BOH equations and for low enough energies, the following single equation:[32]

$$\frac{1}{2m}\left(\frac{\hbar}{i}\frac{\partial}{\partial x} + \hbar\tau_{x12}(x)\right)^2 \Psi(x) + (u_1(x) - E)\Psi(x) = 0 \quad (6)$$

It is important to emphasize that the solution of Eq. (6) is reliable for all the conditions/situations envisaged in the present study. (The relevance of this approximation was tested extensively in Ref. 33 – see Tables (II)-(V)). The similarity between eq. (4) and (6) suggests that $\tau_{12}(x)$ is akin to the friction parameter $\beta$ which is known to be responsible for the classical dissipative force (or friction force).[26,27] Therefore, it follows that

$$\hbar\tau_{x12}(x|s) = \beta(x)x \Rightarrow \beta(x) = \frac{\hbar\tau_{x12}(x|s)}{x}. \quad (7)$$

Such a relation holds for other Cartesian coordinates $y$ and $z$ also, where applicable. Here, it must be emphasized that the NACT is a purely quantum mechanical entity and is being related to classical friction and used in solving the classical equation of motion for a molecular system.

It is worth emphasizing that the adiabatic BOH equation (eq. (6)) does not conserve energy. There is one issue which plays in favor of $\tau_{12}$ being associated with the friction process. That is, during the slowing down process it releases energy as heat — in the present case the energy would be released as light (or, better, as photons), as envisaged in the production of HeH$^+$ in (He$^+$, H$_2$) collisions (see Ref. [34] and discussion in Ref. [27]). One might argue that this should be described quantum mechanically and the rate of dipolar emission should be estimated. Unfortunately, such an approach would take us beyond the BOH treatment and is beyond the scope of present undertaking.

The dissipation process is enhanced once three or more quasi-ions are in relative proximity because, in such a case, the NACTs attain, at certain regions in configuration space, infinite values. As a result, the fast-moving quasi-ions are continuously slowed down and in this way are forced to lose their excessive energy at various rates so that the electrostatic forces may, afterwards, takeover.

In an earlier study,[26] we had examined the possibility of the NACTs acting as a friction force slowing down the dynamics of ($H_2$, $H^+$) collisions in $C_{2v}$ geometry and along parallel axes close to the $C_{2v}$ geometry. For practical reasons, $H_2$ was taken to be in its equilibrium geometry. For the initial kinetic energy (KE) ranging from 0.35 eV to 500 eV, it was shown that friction could indeed slow down the dynamics and bring the molecular system to a standstill (KE = 0). A similar study[27] was carried out for (He, $H_2^+$) collisions in collinear geometry and in $C_{2v}$ geometry for a similar energy range and it was shown that the NACT terms could indeed act as friction forces and slow down the dynamics to make KE = 0.

In the present study, we have computed the ab initio PES for the ground and the first excited electronic state of $H_3^+$ along with the NACTs in three dimensions and solved the classical equations of motion by including friction terms equivalent to the NACTs. It is shown that the ($D^+$, $H_2$) dynamics slows down enough to reach the bottom of the potential well. Therefore, we argue that a classical mechanical study including the NACTs would lead to the formation of $DH_2^+$ in its equilibrium geometry.

In case of $H_3^+$, a wide variety of nonadiabatic interactions is spread over the entire nuclear configuration space. One can observe accidental seams between the ground ($1^1A'$) and the first excited ($2^1A'$) states of $H_3^+$ at the $C_{2v}$ (isosceles) as well as $C_s$ (acute) geometries over the global PESs, whereas the CIs between the first ($2^1A'$) and the second ($3^1A'$) excited states are located at $D_{3h}$ (equilateral) nuclear configurations.[35] Therefore, it is necessary to extend the classical dynamical calculations over the global three-dimensional (3D) PES (rather than the one-dimensional potential energy curve investigated in ref. [26]) of $H_3^+$ to compute the trapping percentage in the global minimum (-9.31 eV relative to isolated D, H and $H^+$) as $DH_2^+$ species and other reaction attributes (reaction probabilities and cross sections) of various processes resulting from $D^+ + H_2$ collisions.

Several experiments, like guided beam,[36-38] drift tube,[39] flow tube,[40] merged beam[41-42] and ion-cyclotron resonance[43] have been carried out in the past to determine the differential/integral cross sections (ICSs) and/or rate coefficients for $D^+(H^+) + H_2(D_2/DH)$ reactions. Adhikari and coworkers have computed those quantities for various charge transfer and non-transfer processes of $D^+ + H_2$ [46] and $H + H_2^+$ [47] reactions employing Fully Close Coupled

(FCC) 3D wave packet approach[44-47] over Beyond Born-Oppenheimer (BBO)[35] based diabatic PESs of the lowest three electronic states of $H_3^+$ ($1^1A'$, $2^1A'$ and $3^1A'$).

Over the ground adiabatic PES of $H_3^+$, various channels of $D^+ + H_2$ reaction are defined by triatomic side lengths ($R_{H^1}$-$R_{H^2}$, $R_{H^1}$-$R_D$ and $R_{H^2}$-$R_D$, where two H atoms are numbered as $H^1$ and $H^2$) as follows:

- Creation of triatomic species (trapping in the minimum of the ground electronic state PES):
  $D^+ + H_2 \rightarrow DH_2^+$ ($R_{H^1}$-$R_{H^2} \approx R_{H^1}$-$R_D \approx R_{H^2}$-$R_D \approx 0.873$ Å)

- Non-reactive non-charge transfer (NRNCT):
  $D^+ + H_2 \rightarrow D^+ + H_2$ ($R_{H^1}$-$R_D > 3$ Å and $R_{H^2}$-$R_D > 3$ Å)

- Reactive non-charge transfer (RNCT):
  $D^+ + H_2 \rightarrow H^+ + DH$ ($R_{H^1}$-$R_D$ or $R_{H^2}$-$R_D > 3$ Å and $R_{H^1}$-$R_{H^2} > 3$ Å)

For the present case, classical trajectory calculations have been carried out over the ground adiabatic PES of $H_3^+$ to compute various reaction attributes of NRNCT, RNCT, and trapping processes.

In this work, nine Cartesian coordinates of $(H, H, D)^+$ and their associated momenta are initialized over multi-reference-configuration interaction (MRCI) based ground adiabatic PES[35] of $H_3^+$ [$1^1A'$, asymptotically: $H_2$ ($^1\Sigma_g^+$) + $D^+$], where the diatom ($H_2$) and the third atom ($D^+$) can approach each other from any direction. Thousands of classical trajectories are computed over the global PES including the NACTs as dissipative friction terms in the classical equations of motion (EOMs). Finally, the percentage of trapped trajectories as well as the effect of friction on various reaction attributes like reaction probabilities and cross sections are evaluated for NRNCT and RNCT processes.

# 3. Initialization, Propagation and Analysis

In this single-surface classical trajectory calculation, the initial positions and momenta of the triatomic reactive system, $D^+ + H_2$ are chosen in Jacobi coordinates by the rotating Morse oscillator approach of Porter et al.[48] and then, those Jacobi coordinates are transformed into the Cartesian counterparts of individual atoms. For the sake of clarity, the two H atoms and the third D atom are numbered 1, 2 and 3, respectively. Once those nine Cartesian coordinates ($x_1$, $y_1$, $z_1$, $x_2$, $y_2$, $z_2$, $x_3$, $y_3$, $z_3$) and their associated momenta ($p_{x_1}$, $p_{y_1}$, $p_{z_1}$, $p_{x_2}$, $p_{y_2}$, $p_{z_2}$, $p_{x_3}$, $p_{y_3}$, $p_{z_3}$) are initialized, classical EOMs are solved over the ground electronic (adiabatic) PES of $H_3^+$ in presence of the NACTs between the ground and the first excited states acting as frictional force.

Thus, the classical EOMs for nine coordinates ($x_k$, $y_k$, $z_k$, $k = 1,2,3$) and their associated momenta ($p_{x_k}$, $p_{y_k}$, $p_{z_k}$) for three atoms (H, H, D) take the following forms:

$$\dot{p}_{x_k} + \frac{\hbar}{mx_k}\left(\tau_{x_k}^{12} p_{x_k}\right) + \frac{dV(x_k,y_k,z_k)}{dx_k} = 0$$

$$\dot{p}_{y_k} + \frac{\hbar}{my_k}\left(\tau_{y_k}^{12} p_{y_k}\right) + \frac{dV(x_k,y_k,z_k)}{dy_k} = 0$$

$$\dot{p}_{z_k} + \frac{\hbar}{mz_k}\left(\tau_{z_k}^{12} p_{z_k}\right) + \frac{dV(x_k,y_k,z_k)}{dz_k} = 0 \qquad (8)$$

$$\dot{x}_k = \frac{p_{x_k}}{m}$$

$$\dot{y}_k = \frac{p_{y_k}}{m}$$

$$\dot{z}_k = \frac{p_{z_k}}{m} \qquad (9)$$

Since the friction term possesses Cartesian coordinates in the denominator, one or more than one Cartesian coordinate may take zero value(s) during the trajectory. Therefore, there is a possibility of encountering a singularity in the close vicinity of those zero values of coordinates leading to an artificial singularity in the friction term ($\beta$). To prevent such an artefact, the single surface classical trajectory calculations are carried out using a cutoff value (0.05 Å) in the Cartesian coordinate close to zero so that the corresponding NACTs take a terminal value $\hbar\tau_{Q_k}^{12}/mQ_t$, ($Q_k = x_k, y_k, z_k$, $k = 1, 2, 3$), if the modulus of the coordinate $Q_k$ is less than or equal to the cutoff value $Q_t$ (= 0.05 Å).

When the quantity $\sqrt{r^2 + R^2}$ ($r$ and $R$ are the Jacobi coordinates) exceeds a predefined value ($C_{max}$, for the present case ~ 6 Å), the propagation of trajectories is terminated, and the Cartesian coordinates are transformed back to their Jacobi counterparts for final analysis. For the trajectories that are not trapped, but leading to either NRNCT or RNCT products, the rovibrational states ($v'$, $j'$) of the final species (reactant/product) are determined by the rotating Morse oscillator equations.[48] The rovibrationally resolved [$\sigma^I_{v',j'}(E_{col})$], rotationally summed vibrationally resolved [$\sigma^I_{v'}(E_{col})$] and total integral cross sections (ICSs) are computed following the standard procedure.[9,20]

For the trapped triatomic species, $DH_2^+$, it is not possible to define its orbital angular momentum $l_{v'j'}$, since the kinetic energy becomes nearly zero and the total energy is approximately equal to the minimum of the potential energy for that specific trajectory. Therefore, we compute the trapping cross section of $DH_2^+$ in an alternative way by taking the difference of the ICSs of single surface classical trajectory calculations with and without friction for NRNCT and RNCT processes. Finally, those individual components are added up to compute the total ICS for the trapping phenomenon.

## 4. Results and Discussion

We had computed the global adiabatic PES of the ground ($1^1A'$) electronic state and the NACTs between the ground ($1^1A'$) and the first excited ($2^1A'$) electronic states of $H_3^+$,[35] by employing multi-reference configuration interaction (MRCI) and numerical finite difference (DDR) methodologies, respectively, as implemented in MOLPRO[49] quantum chemistry software.

In the present single-surface classical calculation of $D^+ + H_2$ reaction within the collision energy ($E_{col}$) range 1.7 eV to 2.5 eV, the trajectories are initiated from $v = 0$, $j = 0$ state of $H_2$ in the ground electronic state and are propagated classically over the global adiabatic PES including the NACTs as frictional force in the classical EOMs.

To follow the slowing down of the three atomic species ($D^+$ and $H_2$) and then, getting trapped into the deep potential well (~ -9.31 eV) of the $D^+$ and $H_2$ system, the variation of kinetic, potential, and total energies is presented as a function of the elapsed time (up to 25000 fs) for a trapped trajectory at 1.7 eV in Figure 1. It can be seen that both the kinetic and potential energies show an oscillatory behavior initially, but gradually decrease approaching zero and the global minimum of $H_3^+$, respectively. The total energy, on the other hand, exhibits a decaying behavior till the termination of the trajectory (25000 fs). The magnitude of the potential ($V_{min}^T$) vis-á-vis the total energy is in the range of -8 to -9 eV at the point of trapping, which is close to the potential minimum ($V_{min}$) of $H_3^+$ (~ -9.31 eV). A careful examination of the bond distances between atoms reveals that the side lengths are around 0.7-1.4 Å for the trapped trajectory, close to the triatomic side length ($R_{min}$) at the global minimum of $H_3^+$ (0.873 Å). Clearly, the triatomic species $DH_2^+$ is formed due to the NACTs acting as friction.

Reaction probabilities were calculated by computing several thousand single-surface classical trajectories for different collision energies on the ground adiabatic PES with and without friction. When friction was NOT included, classical trajectory-based reaction probabilities for both NRNCT and RNCT processes were almost invariant with respect to the collision energy and the total probability was unity. When friction was included, the probability for trapping decreased with an increase in the collision energy, while probabilities for NRNCT and RNCT processes showed a slightly increasing trend. The combined probability for NRNCT and RNCT processes is less than one due to the trapping as $DH_2^+$, but the total probability of all the three processes (trapping, NRNCT and RNCT) remains one for each collision energy.

In the present work, single-surface classical trajectory-based ICSs (see Figure 2) were calculated for NRNCT and RNCT processes of $D^+$ + $H_2$ collision and compared with the quantum results obtained from a Fully Close Coupled (FCC) 3D wave packet approach[46]. For both the processes (see Figures 2a and 2b), classical trajectory-based cross-section values without friction showed good agreement with the quantum results[46], but inclusion of friction terms in the EOMs led to a lowering of the ICS profiles compared to the quantum ones. It is necessary to emphasize that adiabatic and diabatic representations of the same ab initio PES of $H_3^+$ were employed for classical trajectory calculation and quantum dynamics, respectively.

## 5. Conclusion

The inclusion of NACTs as frictional force in classical EOMs slows down the motion of the triatomic species $(D, H, H)^+$ leading to trapping of some of the trajectories near the potential energy minimum of $DH_2^+$ species (~ 9.31 eV). The formation of the triatomic species is evident from the trapping percentage, which is close to 10 – 18 % and gradually decreases with an increase in the collision energy in the range 1.7 eV - 2.5 eV. The effect of the dissipative frictional force is also prominent in the ICS profiles. In presence of friction, the classical trajectory-based ICSs for NRNCT and RNCT processes are lower than the values obtained from the classical dynamics without friction. The differences between those two curves (with and without friction) were used to compute the trapping cross section for $DH_2^+$ (see Figure 2c). It is worthwhile to mention that the cross-section profiles for all the processes (NRNCT, RNCT and trapping) exhibit a decaying trend with an increase in the collision energy.

In summary, the present investigation confirms the feasibility of formation of $DH_2^+$ on the ground electronic state PES due to the NACTs acting as friction. Details of the methodology adopted will be published subsequently.

Ideally, one would have liked to compare the estimated cross section values for $DH_2^+$ formation with the experimentally determined cross section values. Unfortunately, such experimental results are not available till this date. We hope that our study would provide an impetus for such an undertaking.

**Declaration of competing interest**

The authors declare that there are no known conflicts of interest.

**Acknowledgement**

**Funding:** Soumya Mukherjee and Satyam Ravi acknowledge IACS for the research fellowship. Satyam Ravi also thanks the VIT Bhopal University for the seed grant, SupPort for Excellence in Academic Research (SPEAR) (File No: VITB/SEEDGRANT/2022/08). Satrajit Adhikari is thankful to SERB for research funding through Project No.: CRG/2019/000793.

**References**

[1] Baer M., Adiabatic and diabatic representations for atom-molecule collisions: Treatment of the collinear arrangement, *Chem. Phys. Lett*. **1975**, *35*, 112-118.

[2] Kuppermann A.; Abrol R., Quantum reaction dynamics for multiple electronic states, *Adv. Chem. Phys*. **2002**, *124*, 283.

[3] Englman R.; Yahalom A., Complex States of Simple Molecular Systems, *Adv. Chem. Phys*. **2002**, *124*, 197.

[4] Yahalom A., Advances in Classical Field Theory, (Bentham eBooks eISBN: 978-1-60805-195-3, (**2011**), Chap. 9.

[5] Bene E.; Vertesi T.; Englman R., Reciprocity in the degeneracies of some tetra-atomic molecular ions, *J. Chem. Phys.* **2011**, *135*, 084101.

[6] Arasaki Y.; Scheit S.; Takatsuka K., Induced photoemission from nonadiabatic dynamics assisted by dynamical Stark effect, *J. Chem. Phys.* **2013**, *138*, 161103.

[7] Sathyamurthy N.; Mahapatra S., Time-dependent quantum mechanical wave packet dynamics, *Phys. Chem. Chem. Phys.* **2021**, *23*, 7586.

[8] Rivlin T,; Pollak E., Nonadiabatic Couplings Can Speed Up Quantum Tunneling Transition Path Times, *J. Phys. Chem. Lett.* **2022**, *13*, 10558−10566.

[9] Porter R. N.; Raff L. M, Dynamics of Molecular Collisions, Part B, edited by W. H. Miller (Plenum, New York, **1976**).

[10] Bjerre A.; Nikitin E. E., Energy transfer in collisions of excited sodium atoms with nitrogen molecules, *Chem. Phys. Lett.* **1967**, *1,* 179.

[11] Preston R. K.; Tully J. C., Effects of Surface Crossing in Chemical Reactions: The $H_3^+$ System, *J. Chem. Phys.* **1971**, *54*, 4297.

[12] Tully J. C.; Preston R. K., Trajectory Surface Hopping Approach to Nonadiabatic Molecular Collisions: The Reaction of $H^+$ with $D_2$, *J. Chem. Phys.* **1971**, *55*, 562.

[13] Chapman S., The Classical Trajectory-Surface-Hopping Approach to Charge-Transfer Processes, *Adv. Chem. Phys.* **1992**, *82, part 2,* 423.

[14] Gianturco F. A.; Schneider F., Model Potential Energy Surfaces for Inelastic and Charge-Transfer Processes in Ion-Molecule Collisions, *Adv. Chem. Phys.* **1992**, *82, part 2,* 135.

[15] Miller W. H., Classical-limit quantum mechanics and the theory of molecular collisions, *Adv. Chem. Phys.* **1975**, *25*, 69.

[16] Meyer H-D.; Miller W. H., A classical analog for electronic degrees of freedom in nonadiabatic collision processes, *J. Chem. Phys.* **1979**, *70*, 3214.

[17] Child M. S., Theory of Chemical Reaction Dynamics, Baer M., Ed., CRC Press, Boca Raton, **1985**, Vol. III, p. 279.

[18] Maiti B.; Schatz G. C., Theoretical studies of intersystem crossing effects in the $O(^3P,^1D)$ + $H_2$ Reaction, *J. Chem. Phys.* **2003**, *119*, 12360.

[19] Mahata P.; Rauta A. K.; Maiti B., Trajectory surface-hopping study of 1-pyrazoline photodissociation dynamics, *J. Chem. Phys.* **2022**, *157*, 194302.

[20] Mukherjee S.; Hazra S.; Ghosh S.; Mukherjee S.; Adhikari S., Trajectory Surface Hopping vs. Quantum Scattering Calculations on $D^+ + H_2$ and $H + H_2^+$ Reactions using Ab Initio Surfaces and Couplings, *Chem. Phys.* **2022**, *560*, 111588.

[21] Baer M., Beyond Born Oppenheimer; Electronic non-Adiabatic coupling Terms and Conical Intersections, Wiley & Sons Inc, Hoboken N.J. **2006**; (a) Sect 5.1.1; (b) Sect 2.1.1;(c) Sect. 6.2.1.

[22] Mukherjee B.; Naskar K.; Mukherjee S.; Ghosh S.; Sahoo T.; Adhikari S., Beyond Born–Oppenheimer theory for spectroscopic and scattering processes, *Int. Rev. Phys. Chem.* **2019**, *38,* 287.

[23] Born M.; Oppenheimer J. R., On the Quantum Theory of Molecules, *Ann. Phys. (Leipzig),* **1927**, *84*, 457.

[24] Born M., Festschrift Goett. *Nach. Math. Phys.* **1951**, *K1*, 1.

[25] Born M.; Huang K., Dynamical Theory of Crystal Lattices, Oxford University, New York **1954**.

[26] Ravi S.; Mukherjee S.; Mukherjee B.; Adhikari S.; Sathyamurthy N.; Baer M., Non-

adiabatic coupling as a frictional force in the formation of $H_3^+$: a model dynamical study, *Eur. Phys. J. D* **2020**, *74*, 239.

[27] Ravi S.; Mukherjee S.; Mukherjee B.; Adhikari S.; Sathyamurthy N.; Baer M., Non-adiabatic coupling as a frictional force in (He, H, H)$^+$ dynamics and the formation of $HeH_2^+$, *Mol. Phys.* **2021**, *119*, e1811907.

[28] Kramers H. A., Brownian Motion in a Field of Force and the Diffusion Model of Chemical Reactions. *Physica* **1940**, *7*, 284.

[29] Das D.; Mukhopadhyay S., Molecular origin of internal friction in intrinsically disordered proteins, *Acc. Chem. Res.* **2022**, *55*, 3470.

[30] Goldstein H., Classical Mechanics, Addison-Wesley Publishing Company, Cambridge, Mass, **1956**.

[31] Landau L. D.; Lifshitz E. M., Mechanics, Pergamon Press, New York, **1997**.

[32] Baer M.; Englman R., A modified Born-Oppenheimer equation: application to conical intersections and other types of singularities, *Chem. Phys. Lett.* **1997**, *265*, 205.

[33] Baer R.; Charutz D. M.; Kosloff R.; Baer M., A Study of Conical Intersection Effects on Scattering Processes: The Validity of Adiabatic Single-Surface Approximations within a Quasi-Jahn-Teller Model, *J. Chem. Phys.* **1996**, *105*, 9141.

[34] Güsten R.; Wiesemeyer H.; Neufeld D.; Menten K. M.; Graf U. U.; Jacobs K.; Klein B.; Ricken O.; Risacher C.; J. Stutzki, Astrophysical detection of the helium hydride ion $HeH^+$, *Nature,* **2019**, *568*, 357.

[35] Mukherjee S.; Mukhopadhyay D.; Adhikari S., Conical intersections and diabatic potential energy surfaces for the three lowest electronic singlet states of $H_3^+$, *J. Chem. Phys.* **2014**, *141*, 204306.

[36] Ochs G.; Teloy E., Integral cross sections for reactions of $H^+$ with $D_2$: New measurements, *J. Chem. Phys.* **1974**, *61*, 4930-4931.

[37] Schlier Ch.; Nowotny U.; Teloy E., Proton-Hydrogen collisions, *Chem. Phys.* **1987**, *111*, 401-408.

[38] Niedner G.; Noll M.; Toennies J. P., Observation of vibrationally resolved charge transfer in $H^+ + H_2$ at $E_{CM} = 20$ eV, *J. Chem. Phys.* **1987**¸ *87*, 2685.

[39] Fehsenfeld F. C.; Albritton D. L.; Bush Y. A.; Fournier P. G.; Govers T. R.; Fournier J., Ion-atom interchange reactions using isotopic species, *J. Chem. Phys.* **1974**, *61*, 2150-

2155.

[40] Henchman M. J.; Adams N. G.; Smith D., The isotope exchange reactions $H^++D_2 \rightleftarrows HD+D^+$ and $D^++H_2 \rightleftarrows HD+H^+$ in the temperature range 200–300 K, *J. Chem. Phys.* **1981**, *75*, 1201-1206.

[41] Baer M.; Niedner G.; Toennies J. P., A three-dimensional quantum mechanical study of vibrationally resolved charge transfer processes in $H^++H_2$ at $E_{cm}$ = 20 eV, *J. Chem. Phys.* **1989**, *91*, 4169.

[42] Baer M.; Niedner G.; Toennies J. P., A quantum-mechanical study of charge transfer steric factors for the three isotopic systems: $H^+ + H_2, D_2, HD$, *Chem. Phys. Lett.* **1990**, *167*, 269.

[43] Karpas Z.; Anicich V.; Huntress W. T., An ion cyclotron resonance study of reactions of ions with hydrogen atoms, *J. Chem. Phys.* **1979**¸ *70*, 2877-2881.

[44] Sahoo T.; Ghosh S.; Adhikari S.; Sharma R.; Varandas A. J. C., Coupled 3D Time-Dependent Wave-Packet Approach in Hyperspherical Coordinates: Application to the Adiabatic Singlet-State ($1^1A'$) $D^+ + H_2$ Reaction, *J. Phys. Chem. A* **2014**¸ *118*, 4837-4850.

[45] Sahoo T.; Ghosh S.; Adhikari S.; Sharma R.; Varandas A. J. C., Low-temperature $D^+ + H_2$ reaction: A time-dependent coupled wave-packet study in hyperspherical coordinates, *J. Chem. Phys.* **2015**¸ *142*, 024304.

[46] Ghosh S.; Mukherjee S.; Mukherjee B.; Mandal S.; Sharma R.; Chaudhury P.; Adhikari S., Beyond Born-Oppenheimer theory for ab initio constructed diabatic potential energy surfaces of singlet $H_3^+$ to study reaction dynamics using coupled 3D time-dependent wave-packet approach, *J. Chem. Phys.* **2017**¸ *147*, 074105.

[47] Ghosh S.; Sahoo T.; Baer M.; Adhikari S., Charge Transfer Processes for $H + H_2^+$ Reaction Employing Coupled 3D Wavepacket Approach on Beyond Born–Oppenheimer Based Ab Initio Constructed Diabatic Potential Energy Surfaces, *J. Phys. Chem. A* **2021**¸ *125*, 731-745.

[48] Porter R. N.; Raff L. M.; Miller W. H., Quasiclassical selection of initial coordinates and momenta for a rotating Morse oscillator, *J. Chem. Phys.* **1975**, *63*, 2214.

[49] Werner H. –J.; Knowles P. J.; Knizia G.; Manby F. R., *et al.* MOLPRO, version 2012.1, a package of *ab initio* programs, see http://www.molpro.net

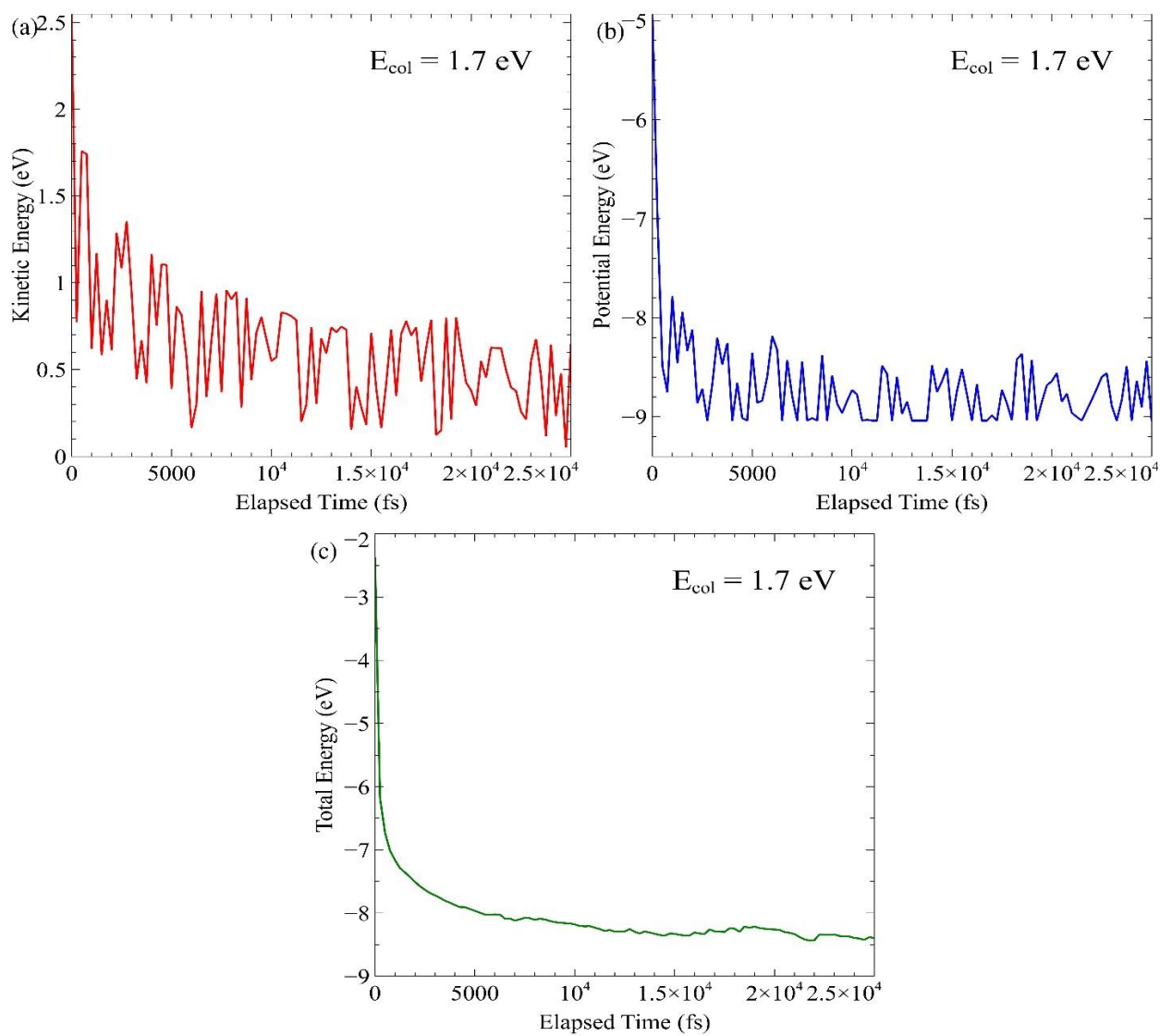

Figure 1. Variation of (a) kinetic energy, (b) potential energy and (c) total energy with time at $E_{col}$ = 1.7 eV (cutoff in coordinate = 0.05 Å). The trajectory is trapped where the potential energy ($V_{min}^T$) is -9.039 eV and total energy is -8.395 eV.

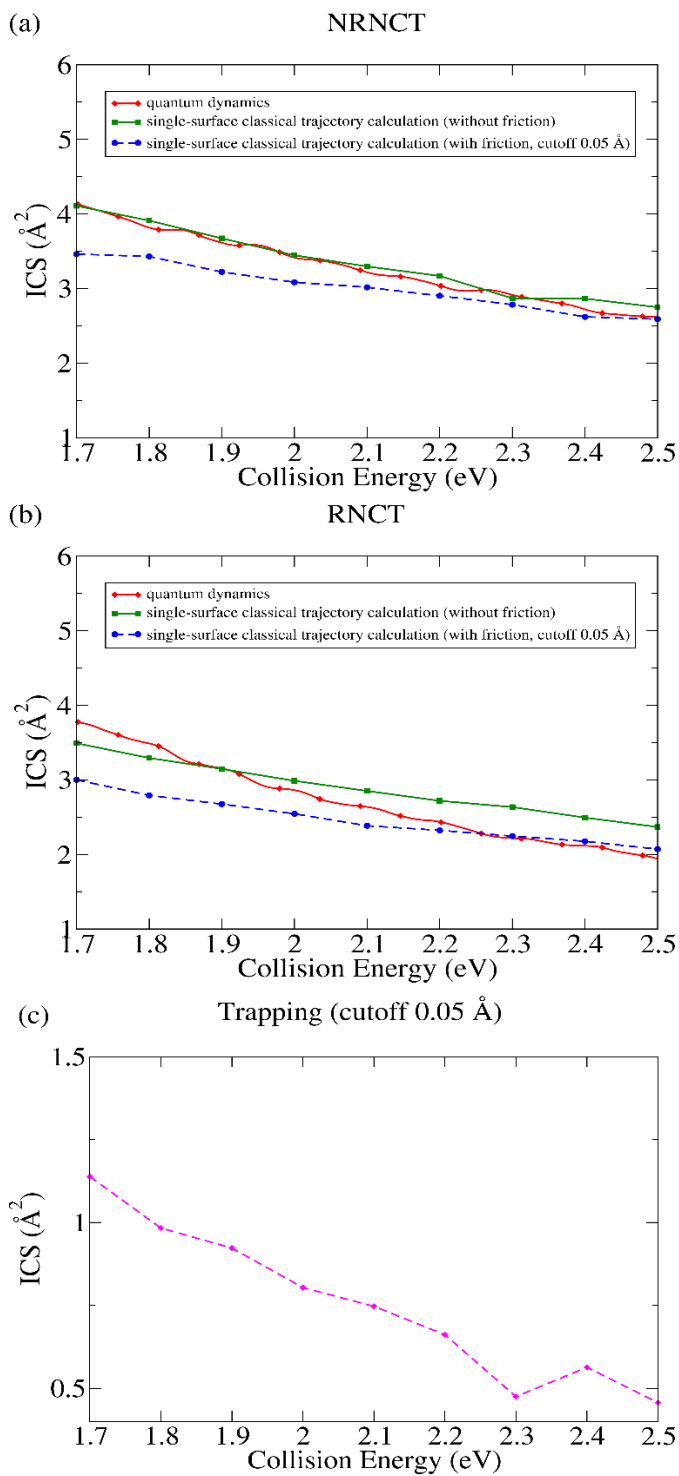

Figure 2. Integral cross section values plotted as a function of collision energy (1.7 eV to 2.5 eV) for (a) NRNCT and (b) RNCT processes, obtained by single-surface quasi-classical trajectory calculations, with and without fraction and compared with the quantum results [46] (no friction) on the same surface. Included in panel (c) are the ICS values for the trapping process.